\documentclass[aps,prl,twocolumn,showpacs,superscriptaddress,preprintnumbers,amsmath,amssymb]{revtex4}
\usepackage{amsmath}
\usepackage{graphicx}% Include figure files
\usepackage{dcolumn}% Align table columns on decimal point
\usepackage{bm}% bold math
\usepackage{float}
\usepackage{xcolor}
\begin{document}
\preprint{\textit{Preprint}}
%\preprint{APS/123-QED}
%\begin{CJK*}{GB}{gbsn}  %%%%\begin{CJK*}{GBK}{song}
\title{High-Harmonic Generation and Spin-Orbit Interaction of Light in a Relativistic Oscillating Window}
% Force line breaks with \\
\author{Longqing Yi}
\thanks{lqyi@sjtu.edu.cn}
\affiliation{Department of Physics, Chalmers University of Technology, 41296 Gothenburg, Sweden}

\date{\today}% It is always \today, today,

%\keywords{Keyword1, Keyword2, Keyword3}

\begin{abstract}
When a high power laser beam irradiates a small aperture on a solid foil target, the strong laser field drives surface plasma oscillation at the periphery of this aperture, which acts as a ``relativistic oscillating window". The diffracted light that travels though such an aperture contains high-harmonics of the fundamental laser frequency. When the driving laser beam is circularly polarised, the high-harmonic generation (HHG) process facilitates a conversion of the spin angular momentum of the fundamental light into the intrinsic orbital angular momentum of the harmonics. By means of theoretical modeling and fully 3D particle-in-cell simulations, it is shown the harmonic beams of order $n$ are optical vortices with topological charge $|l| = n-1$, and a power-law spectrum $I_n\propto n^{-3.5}$ is produced for sufficiently intense laser beams, where $I_n$ is the intensity of the $n$th harmonic. This work opens up a new realm of possibilities for producing intense extreme ultraviolet vortices, and diffraction-based HHG studies at relativistic intensities.
\end{abstract}

\maketitle

Light carries angular momentum as spin and orbital components. The spin angular momentum (SAM) is associated with right or left circular polarisation ($\pm\hbar$ per photon), and the orbital angular momentum (OAM) is carried by light beams with helical phase fronts $\exp(il\phi)$ ($l\hbar$ per photon), also known as optical vortices, where $l$ is the topological charge and $\phi$ is the azimuthal angle \cite{Allen1992}.
The spin-orbit interaction of light refers to phenomena in which the spin affects the orbital degrees of freedom \cite{Bliokh2015a}, such as spin-Hall effects \cite{Onoda2004,Hosten2008}. %spin-dependent effects in nonparaxial fields\cite{Bliokh2010} and evanescent waves \cite{Bliokh2015b}.
Recently, interest in spin-orbit interaction has surged, as it not only gives physical insights into the behaviour of polarised light at sub-wavelength scales,
%which is essential in nano-optics and photonics,
%In addition, spin-orbit angular momentum conversion is
but also provides an important approach for producing optical vortices in the extreme ultraviolet (XUV) regime \cite{Dorney2019,Zurch2012,Garcia2013,Gariepy2014,Gauthier2017}, that have a rich variety of applications in optical communication \cite{Wang2012,Gibson2004}, biophotonics \cite{Willig2006}, and optical trapping \cite{ONeil2002}.

%The production of optical vortices with such methods mostly rely on high-harmonic generation (HHG) in laser-atom interactions, driven by a moderately intense ($\sim10^{14}$W/cm$^{2}$) beam \cite{Dorney2019,Zurch2012,Garcia2013,Gariepy2014,Gauthier2017}. The resulting extreme ultraviolet (XUV) vortices are of particular interest for monitoring and manipulating the SAM and OAM of light-matter interactions on the atomic scale, as well as for applications such as nonlinear optics \cite{Patchkovskii2012,Veenendaal2007} and superresolution microscopy \cite{Hell1994}.
Owing to the remarkable progresses in high-power lasers \cite{CPA}, such advanced light sources open up new possibilities in the relativistic regime ($>10^{18}$W/cm$^{2}$) of light-matter interactions \cite{Mendonca2009,Shi2014,Vieira2016,Leblanc2017,Vieira2018,Wang2020}, and can yield fundamental insights into the spin-orbit and orbit-orbit angular momentum interactions of relativistic light \cite{Zhang2015,Zhang2016,Denoeud2017,Tang2019}.
In particular, intense, ultrafast XUV vortices are of great interest for probing and manipulating the SAM and OAM of light-matter interactions on the atomic scale.
Most of the proposed methods to produce such beams are based on high-harmonic generation (HHG) driven by relativistic vortex laser beams \cite{Vieira2016,Zhang2015,Denoeud2017}, that are not widely available.
Other techniques employ linearly-polarised laser beams interacting with plasma holograms \cite{Leblanc2017}, or circularly polarised (CP) laser pules irradiating a dented target \cite{JWang2019,Li2020}.
However, these approaches rely on the relativistic oscillating mirror (ROM) mechanism \cite{Bulanov1994,Lichters1996,Baeva2006} for producing harmonics, which is suppressed for CP drivers at normal incidence~\cite{Baeva2006,Chen2016}. Therefore it is challenging to generate intense circularly polarised vortex beams that are of particular interest for controlling chiral structures \cite{Toyoda2012,Toyoda2013} and optical manipulation at relativistic intensities \cite{WWang2019}, due to the unique feature of constant ponderomotive force and donut-shaped intensity.

%including driving twisted plasma waves carrying OAM \cite{Mendonca2009,Shi2018,Vieira2018}, with applications in advanced particle accelerators \cite{Wang2020} and light sources \cite{Vieira2016}, as well as revealing fundamental insight into the spin-orbit/orbit-orbit angular momenta interplay of relativistic light \cite{Zhang2015,Zhang2016,Denoeud2017,Tang2019}.

%studying spin-to-orbital angular momentum conversion \cite{Zhao2007}.
%It was reported recently \cite{WWang2019} that such beams can allow for optical manipulation at relativistic intensities, due to the unique feature of constant ponderomotive force and donut-shaped intensity.}

%It was reported recently \cite{Wang2019,Li2020} that XUV vortices can be generated by irradiating a solid foil with a circularly polarised (CP) high-power laser, the SAM of the driver is converted into OAM of the harmonics through the relativistic oscillating mirror (ROM) mechanism \cite{Bulanov1994,Lichters1996,Baeva2006}. This approach is attractive as high-power CP drivers are more accessible.
%However, according to the ROM theory \cite{Baeva2006}, HHG is suppressed for a CP driver at normal incidence. Therefore the mechanism typically produces relatively-weak harmonic intensities \cite{Wang2019}.
%relies crucially on the pre-denting of the target surface by radiation pressure, and

%Here we present, for the first time, a semi-analytical theory of HHG based on light diffraction at relativistic intensities. A new HHG mechanism is identified, which we call relativistic oscillating window (ROW).
In this Letter, we introduce a new HHG mechanism based on light diffraction at relativistic intensities \cite{GI2016NP,GI2016NC,Duff2020}, which we call relativistic oscillating window (ROW). It allows for producing ultra-intense circularly-polarised XUV vortices with a high-power CP laser beam.
We show that when the laser pulse propagates through a small aperture on a thin foil, it drives chiral electron oscillation at the periphery, which results in spin-orbit interaction and HHG in the diffracted light.\\

%that have helical phase fronts with topological charge $|l| = n-1$ for the harmonic order $n$; a universal power-law spectrum $I_n\propto n^{-3.5}$ is produced at the ultra-relativistic limit, where  $I_n$ is the intensity of $n$th harmonic.\\
%$n$ is the harmonic order and

\begin{figure*}[!t]
\centering
\includegraphics[width=0.8\textwidth]{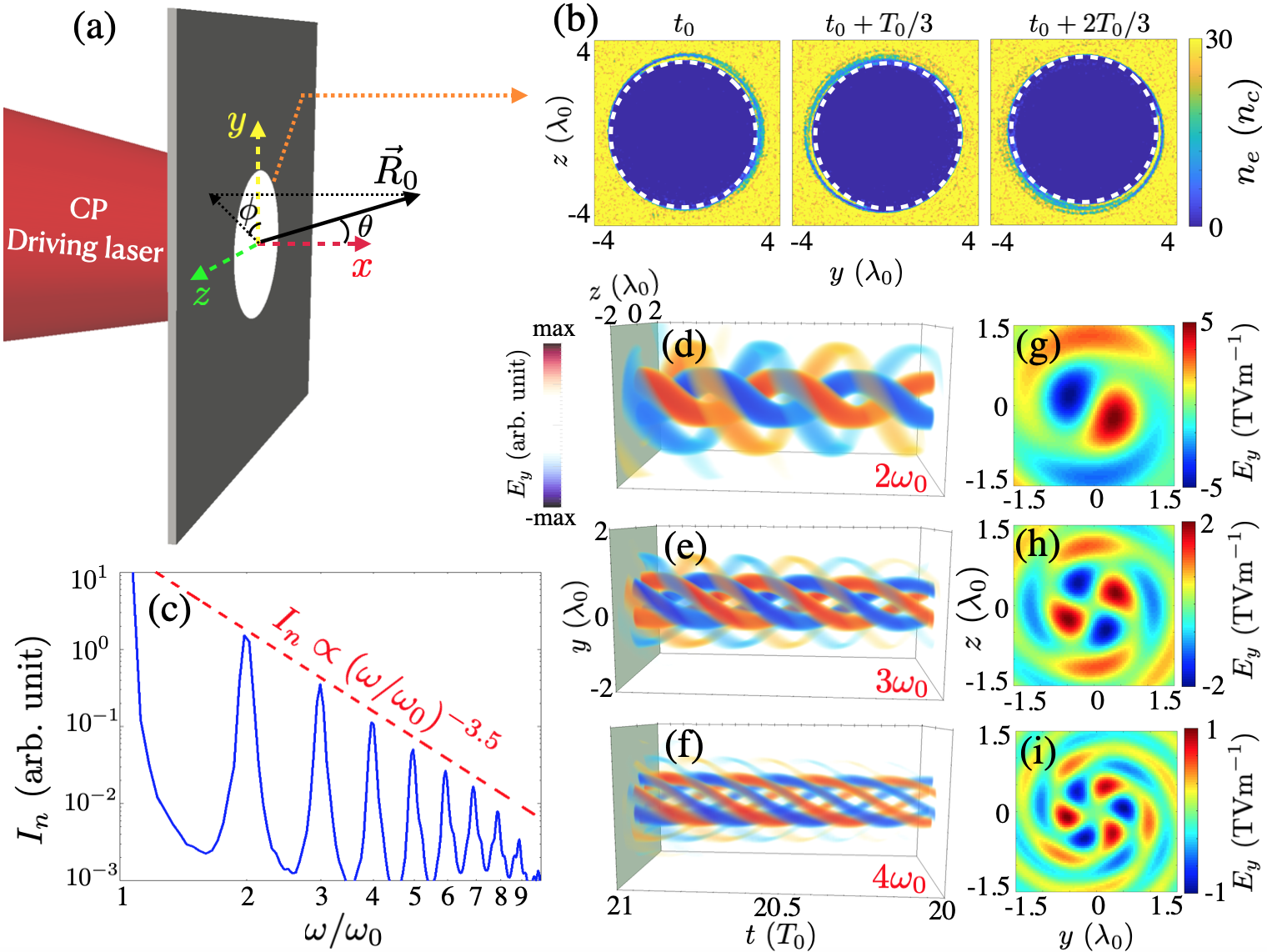}
\caption{(a) An intense CP laser beam is focused on a foil with a small aperture, the laser field drives surface electron oscillation at the periphery, resulting in a dynamical electron density distribution (b). The three snapshots are separated temporally by a third of laser period ($T_0$), from
left to right, and the white dashed lines represent the boundary of a rigid oscillating window. (c) The spectrum of the diffracted light, the red
dashed line represents a fitted power-law spectrum $I_n\propto n^{-3.5}$. (d-f) show the the harmonic fields with frequency $2\omega_0$, $3\omega_0$,
and $4\omega_0$, respectively.
%The coordinate system used in the PIC simulation and the analytical model are presented in (a).
%These fields are obtained from PIC simulations by recording $E_y$ passing through an observational plane placed at $x = x_0 + 10 \mu$m.
The field distributions in the 2D planes marked by dark green colour in (d-f) are shown in (g-i), respectively.}
\vspace{-1pt}
\end{figure*}

We first demonstrate our scheme using 3D particle-in-cell (PIC) simulations with the code {\sc epoch} \cite{Arber2015}. The simulation setup
and the main results are summarised in Fig.~1: a CP laser beam propagates through a small aperture on a thin foil  located at $x_0 = 4~\rm{\mu m}$.
%normally on a thin foil located at $x_0 = 4~\rm{\mu m}$ with a small aperture aligned with the laser beam.
The laser field used in the simulation is $\mathbf{E}_l = (\mathbf{e_y}+i\sigma\mathbf{e_z})E_0\sin^2(\pi t/\tau_0)\exp(ik_0x-i\omega_0 t)$,
$0<t<\tau_0 = 54$ fs,where $\mathbf{e_y}$ ($\mathbf{e_z}$) are the unit vectors in $\mathbf{y}$ ($\mathbf{z}$) direction, $E_0$ is the laser
amplitude, $k_0 = 2\pi/\lambda_0$ the wavenumber, and $\lambda_0 = 1 {\rm \mu m}$ the wavelength. The laser polarisation is controlled
by $\sigma = +1$  and $-1$ for right- and left-handed circular polarisation, respectively. The intensity of the laser beam
is $I_0 \approx 6.9\times10^{19}$ W/cm$^{2}$, corresponding to a normalised laser amplitude of $a_0 \equiv eE_0/m_{\rm{e}}c\omega_0 = 5$,
where $e$, $m_{\rm{e}}$, $c$ and $\omega_0$ denote  the elementary charge, electron mass, vacuum light speed, and the laser frequency, respectively.
%Note that we have assumed the laser focus spot is much greater than the size of the aperture, so that the intensity on the edge of the aperture does not depend on the its radius.
%This is only for the convenience of comparing the PIC results with our model, not crucial for the proposed mechanism to work.
The thin foil target [assumed plastic (CH)] is modeled by a pre-ionised plasma with thickness $L_{\rm f} = 0.25~{\rm \mu m}$, and electron density $n_0 = 30n_{\rm{c}}$, where $n_{\rm{c}} = m_{\rm{e}}\omega_0^2/4\pi e^2 \approx 1.1\times 10^{21}$ cm$^{-3}$ is the critical density. The radius of the aperture is $r_{\rm{A}} = 4.0~{\rm \mu m}$, with a density gradient at the inner boundary $n(r) = n_0\exp[(r-r_{\rm{A}})/h]$ for $r<r_{\rm{A}}$, where $h = 0.2 {\rm \mu m}$ is the scale length. This yields an effective radius $r_0 = 3.3\mu$m, for which $n(r_0) = 1n_{\rm c}$.
The dimensions of the simulation box are $L_x\times L_y\times L_z = 15 {\rm \mu m}\times 16 {\rm \mu m} \times 16 {\rm \mu m}$, sampled by $2400\times320\times320$ cells with fourteen macroparticles for electrons, two for C$^{6+}$ and two for H$^{+}$ per cell.
Mobile ions with real charge-to-mass ratio are used in the PIC simulations. A high-order particle shape function is applied to suppress numerical
self-heating \cite{Arber2015}. An open boundary condition is used in the $\pm x$ direction, while in the $\pm y$ and $\pm z$ directions, a periodic
boundary condition is applied to launch a plane-wave laser pulse. This is justified as the size of the focal spot is assumed to be much larger than the aperture.

%Thus the simulation box captures only the underlying physics near the aperture, which allows for reaching maximum numerical resolution.
%and the algorithm developed by Cowan et al.~\cite{Cowan2013} is used to minimise the numerical dispersion.
%The intensity on the edge of the aperture does not depend on the its radius, thus easier to compare the simulation results with our model.

The intense laser field drives surface electron oscillations at the periphery, which modify the local plasma density as shown in Fig.~1(b).
Since the region with electron density above $n_c$ is reflective to the laser pulse, the transparent area acts as a ``relativistic oscillating window". %For an ultra-relativistic laser $a_0\gg1$, the window oscillation amplitude is $\sim c/\omega_0$, and its period equals to the laser cycle ($T_0  = 3.3$~fs).

Figure~1(c) presents a typical spectrum of the diffracted light, which contains both even and odd orders of harmonics. It has a power-law shape that can be fitted by $I_n\propto n^{-3.5}$. The spectrum is obtained as the Fourier transform of the fields observed at a vertical plane 11~$\mu$m away from the screen, within an opening angle of $\theta = 30^{\circ}$.

Each harmonic with order $n$ is then selected by spectral filtering in the frequency range $[n-0.5,n+0.5]\omega_0$, shown in Figs.~1(d-i).
The spin-orbit interaction of light takes place, all harmonics are optical vortices with $|l| = n-1$.% The sign of $l$ is controlled by the polarisation of the driving laser ($\sigma$), which determines the chirality of the oscillating window, and has a profound impact on the orbital degree of freedom of the harmonics.\\

Note that the underlying physics of the ROW, i.e., the chiral surface electron oscillation on the rim of the window, is a robust process for CP light diffraction at relativistic intensities. The proposed scheme can work at both normal and oblique incidence, and a self-generated aperture can be relied on to overcome the alignment issue (see \textbf{Supplemental Material}).\\

In this work, we restrict ourselves to the case of an intense CP light diffracting through a pre-drilled aperture at close-to-normal incidence.
%As discussed by Baeva et al.~\cite{Baeva2006}, the tangential component of electric field associated with a plasma surface current is negligibly small compared to the laser field $E_0$. Therefore
In the following we consider the diffraction of a monochromatic plane wave $\mathbf{E}(x,y,z,t) = \mathbf{U}(x,y,z)\exp{(-i\omega_0t)}$ through an
oscillating aperture, where $\mathbf{U}(x,y,z)$ satisfies Helmholtz equation $(\nabla^2 + k_0^2)\mathbf{U} = 0$. As the target is overdense, it is
reasonable to assume the tangential components of the electric field vanish everywhere except in the aperture~\cite{Baeva2006}, where they can be
approximated by that of the incoming laser fields. The diffracted field is given by the generalised Kirchhoff integral~\cite{Jackson}:
%\begin{equation}
%\mathbf{U}(x,y,z) = \frac{1}{2\pi}\nabla\times\int_{\rm A}(\mathbf{e_n}\times \mathbf{U})\frac{\exp{(ik_0R)}}{R}ds',
%\label{eq1}
%\end{equation}
\begin{equation}
\begin{split}
\mathbf{E}&_{\rm diff}(x,y,z,t) = \mathbf{U}(x,y,z)\exp(-i\omega_0t) \\
&= \frac{1}{2\pi}\nabla\times\int_{\rm A}(\mathbf{e_n}\times \mathbf{U})\frac{\exp{[ik_0R'-i\omega_0t]}}{R'}ds',
\end{split}
\label{eq1}
\end{equation}
where the integration is only over the aperture, $\mathbf{e_n}$ is the unit vector normal to the screen. The distance between an observer at $(x,y,z)$ and elementary source [$ds'(y',z')$] is $R' = |\mathbf{R}-d\mathbf{R'}|$, measured at retarded time $t' = t - R'/c$. Here $\mathbf{R}$ is the initial distance, and $d\mathbf{R'}(y',z',t')$ denotes the shift of $ds'$ due to the strong laser field.

%$R = \sqrt{(x-x')^2+(y-y')^2+(z-z')^2}$ is the distance from the elementary source $ds'$ at $(x',y',z')$ to the observation point $(x,y,z)$.

We now introduce the ROW model, it assumes that the shape of the aperture does not change (rigid window), such that each $ds'$ is shifted by the same amount of displacement, $d\mathbf{R'}(y',z',t') = d\mathbf{R'}(t')$. This is valid for weakly-relativistic drivers, where the surface electrons are simply shifted antiparallel to the driving laser field, resulting in a harmonic oscillation $d\mathbf{R'}(t') = -(\mathbf{e_y}+i\sigma\mathbf{e_z})\delta r_0\exp(-i\omega_0t')$, where $\delta r_0$ is the amplitude of the oscillation, for which values will be given below. To calculate the diffracted fields, one must solve for the retarded time ($t'$) numerically according to the motion of the source:
\begin{equation}
R'(t') = |\mathbf{R}+(\mathbf{e_y}+i\sigma\mathbf{e_z})\delta r_0\exp[ik_0R'(t')-i\omega_0t]|.
\label{eq2}
\end{equation}

%Note that Eq.~(1) is valid for both circularly and linearly polarised (LP) drivers, however the respective behaviours of $R'(t')$ are distinct. For a LP driver, the window oscillates mostly in the polarisation direction, which produces HHG beams carrying no OAM.
%For the purpose of the present work, we will proceed with CP drivers. At weakly-relativistic intensities, the surface electrons are simply shifted antiparallel to the driving laser field, resulting in a harmonic oscillation $d\mathbf{R'}(t') = -(\mathbf{e_y}+i\sigma\mathbf{e_z})\delta r_0\exp(-i\omega_0t')$, where $\mathbf{e_y}$ ($\mathbf{e_z}$) is the unit vectors in $\mathbf{y}$ ($\mathbf{z}$) direction. The rim of the window is always attached to these oscillating electrons.

\begin{figure}[!b]
\centering
\includegraphics[width=8.5cm]{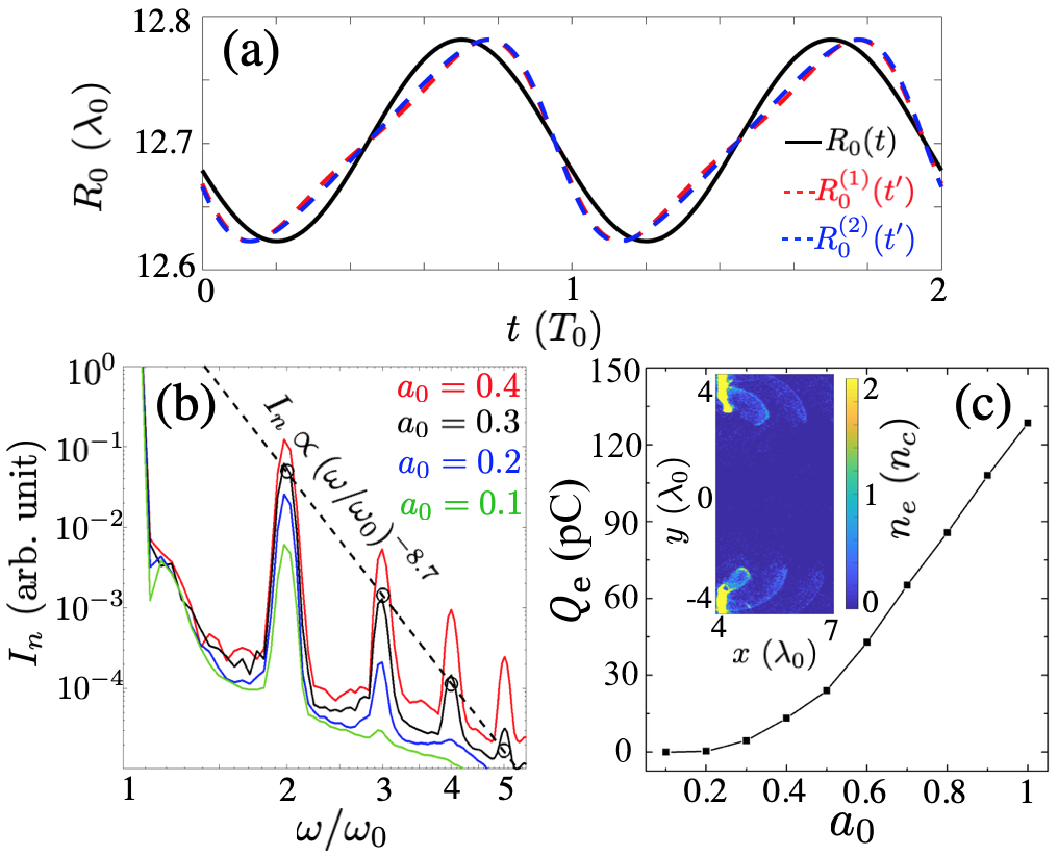}
\caption{(a) The distance between the ROW centre and an observer.
%10~$\mu$m away from the screen, with diffraction angle $\theta = 10^{\circ}$.
The harmonic oscillation of the window is presented by the black curve, and the motion seen by the observer, obtained by solving Eq.~(2), are shown with the red and blue dashed lines, representing the results after one and two iterations, respectively. (b) HHG spectra (solid curves) for drivers with $a_0 = 0.1$ (green), $0.2$ (blue), $0.3$ (black), and $0.4$ (red). The black open circles show the prediction of the ROW model with $\delta r_0 = c/\omega_0$.
(c) The total charge of the escaped electrons is plotted against the laser amplitude $a_0$, the inset shows a typical electron density distribution in $x$-$y$ plane, when SWB occurs.}
%and the black dashed line shows the fitting by $I_n\propto n^{-8.7}$.
%The $E_y$ field distribution of the (b) second, (c) third, and (d) fourth harmonics, predicted by Eq.~(1).}
\label{fig:2}
\end{figure}

However, to explain the spin-orbit interaction, it is sufficient to derive analytically the lowest order of diffracted fields, valid for $a_0\ll1$, seen by a distant, paraxial observer, that satisfies ($R\gg r \gg r_0, \delta r_0$). In this case we have $R'(t') \approx R+\delta r_0\sin(\theta)\exp(ik_0R-i\omega_0t+i\sigma\phi)$, where $\theta = \arctan[r/(x-x_0)]$ and $\phi$ are defined Fig.~1(a).
%The phase is $\Phi = k_0R'-\omega_0t \approx k_0R-\omega_0t+\epsilon\exp(ik_0R-i\omega_0t+i\sigma\phi)$, with $\epsilon = k_0\delta r_0\sin(\theta)\ll1$.
%and the terms that proportional to $\epsilon^2$ and smaller are neglected.
Substituting it into Eq.~(1) and using the Jacobi-Anger identity \cite{JAI}, yields
%\begin{equation}
%%\begin{split}
%\mathbf{E}_{\rm diff} \approx (\mathbf{e_y}+i\sigma\mathbf{e_z})E_0\sum_{n=1}^{\infty} {\rm J}_{n-1}(\epsilon)\frac{-ir_0}{2\pi}\frac{{\rm J}_1[nk_0r_0\sin(\theta)]}{\sin(\theta)}\frac{\exp{[ink_0R_0-in\omega_0t+i(n-1)\sigma\phi]}}{R_0},
%\end{split}
%\label{eq4}
%\end{equation}
\begin{equation}
\begin{split}
\mathbf{E}&_{\rm diff} \approx E_0(\mathbf{e_y}+i\sigma\mathbf{e_z})\sum_{n=1}^{\infty} \frac{-ir_0}{2\pi}\frac{{\rm J}_1[nk_0r_0\sin(\theta)]}{\sin(\theta)}\\
&\times {\rm J}_{n-1}(\epsilon)\frac{\exp{[ink_0R_0-in\omega_0t+i(n-1)\sigma\phi]}}{R_0},
\end{split}
\label{eq3}
\end{equation}
where $\epsilon = k_0\delta r_0\sin(\theta)\ll1$, $R_0 =  \sqrt{(x-x_0)^2+y^2+z^2}$ is the distance measured from the initial centre of the aperture, and ${\rm J}_n$ are the Bessel functions of the first kind. The terms which are proportional to $\epsilon^2$ and smaller are neglected.

Equation~(3) shows that HHG beams have helical phase fronts, with $l = (n-1)\sigma$ for the $n$th harmonic. It agrees well with the findings from PIC simulations. This relation guarantees the conservation of total angular momentum and energy: when $n$ photons at the fundamental frequency are transformed into one photon of $n$th-order harmonic, their SAMs ($n\sigma\hbar$) are converted into $(n-1)\sigma\hbar$ OAM plus $\sigma\hbar$ SAM.\\

In order to examine the ROW model at higher intensities, Eqs.~(1-2) must be solved iteratively. Figure~2(a) presents a typical solution of Eq.~(2)
for the distance between the centre of the window and an observer at $\theta = 30^{\circ}$. It shows that due to the time it takes for the light to
propagate, a harmonic oscillation of the source results in an anharmonic oscillation seen by the observer. This distortion due to retardation is the
dominant mechanism to generate the high harmonics~\cite{Lichters1996}.
%by comparing the HHG spectra and the diffracted field obtained from the model and the PIC simulations
%~\cite{Lichters1996}
% located 11~$\mu$m away from the screen
%, similar to the approach employed by Lichter et. al.~\cite{Lichters1996} to study ROM.

The HHG spectrum is then obtained by Fourier transforming the diffracted field calculated from Eq.~(1). Figure~2(b) shows the spectra for weakly-relativistic drivers. The harmonic intensities increase dramatically with laser $a_0$. In particular, the spectrum for small $a_0$ decays faster than exponentially with $n$, which agrees with Eq.~(3) since ${\rm J}_{n-1}(\epsilon)\sim(\epsilon/2)^{n-1}/(n-1)!$. As $a_0$ grows, the spectrum asymptotically converges to a power-law shape $I_n\propto n^{\alpha}$. This trend can be reproduced by our model as indicated by the open circles in Fig.~2(b).

Equation~(2) suggests the amplitude of the oscillating velocity is $\delta r_0\omega_0$. Thus, substituting $\delta r_0 < c/\omega_0$ into Eq.~(1), one obtains the power-law exponent $\alpha < -8.7$, limited by causality.
However, this is only true for $a_0<0.3$ according to Fig.~2(b), because at higher intensities, the electrons oscillating on the boundary of the aperture may gain enough energy to escape \cite{Naumova2004,Yi2016,Yi2019}, as shown by the inset of Fig.~2(c). Therefore, the rim of the window can no longer be considered to be attached to these electrons, which significantly modifies the dynamics of the ROW.

This is due to surface wave breaking (SWB) \cite{Tian2012}. To quantify when it should be taken into account, in Fig.~2(c) we plot
the total charge of the escaping electrons as a function of the driving laser amplitude $a_0$. A surge of electron emission
is observed for $a_0>0.3$, when the power-law exponents obtained from PIC simulations exceed $-8.7$.\\ %Therefore, the SWB effect is essential to interpret the HHG process in the ultra-relativistic regime.\\
% divided by the total electron number in the skin layer ($N_s\approx 2\pi n_cr_0L_{\rm f}c/\omega_0$),

\begin{figure}[!t]
\centering
\includegraphics[width=8.5cm]{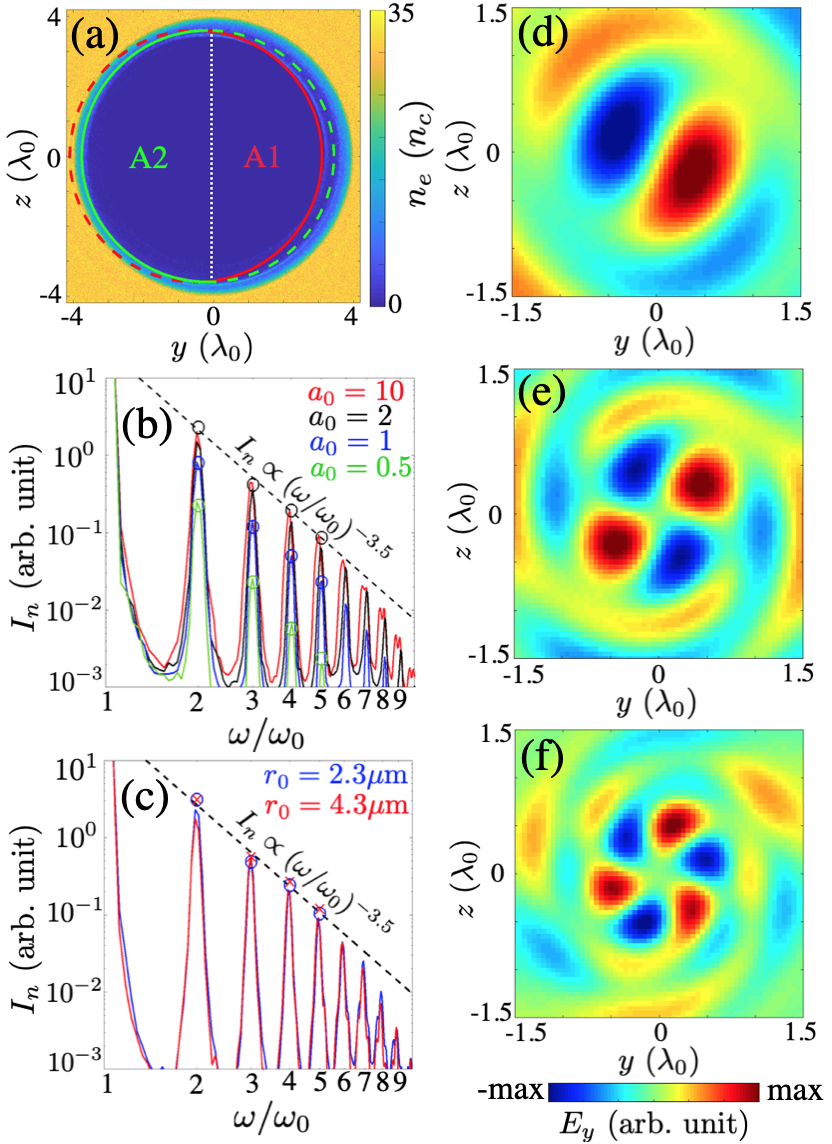}
\caption{(a) The electron density near the aperture when SWB occurs. The window (transparent area) is bounded by the solid curves (red and green). It can be separated into A1 and A2, which are fractions of two other rigid ROWs (red and green circles, consisting of both solid and dashed lines), with different amplitudes $\delta r_{\rm A1}>\delta r_{\rm A2}\approx c/\omega_0$.
(b) HHG spectra from PIC simulations ($r_0 = 3.3\lambda_0$ is fixed) are shown by the solid curves for $a_0 = 0.5$ (green), $1$ (blue), $2$ (black), and $10$ (red). The black open circles are the prediction of ROW model taking into account the SWB effect, with $\delta r_{\rm A1} = 0.25\lambda_0$ (green), $0.4\lambda_0$ (blue), and $0.5\lambda_0$ (black), respectively. (c) The HHG spectra with different radii $r_0 = 2.3 {\rm \mu m}$ and $4.3 {\rm \mu m}$, the PIC simulation data ($a_0 = 5$ is fixed) are shown by the blue and red lines, while the results from the ROW model ($\delta r_{\rm A1} = 0.5\lambda_0$) are presented by the blue open circles and red crosses, respectively. The $E_y$ field of the (d) second, (e) third, and (f) fourth harmonics obtained from the ROW model.}
\label{fig:3}
\end{figure}

We now extend the ROW model to relativistic intensities ($a_0>1$). Figure~3(a) shows a snapshot of typical plasma density distribution near the aperture when SWB occurs. The electrons can now travel far into the aperture when they oscillate inwards, the rim of the window on this side (red solid curve) follows the motion of the electrons for about half of one laser cycle, then it falls back to the original boundary as the electrons are emitted away and transparency is restored. On the other side (green solid curve), when the electrons travel towards the plasma bulk, the displacement remains small.

%Although the electron dynamics in this case inevitably leads to the deformation of the aperture, so that the ``rigid window" assumption is no longer valid. We show that the

The diffracted field can then be calculated by separating the aperture into two parts, A1 and A2. As shown by Fig.~3(a), they are fractions of two rigid ROWs, which oscillate with different amplitudes $\delta r_{\rm A1}>\delta r_{\rm A2}\approx c/\omega_0$. The contributions from each part can then be obtained by integrating Eq.~(1) over the area that satisfies $\mathbf{r}\cdot d\mathbf{R'} \leq 0$ and $\mathbf{r}\cdot d\mathbf{R'} > 0$ for A1 and A2, respectively.
In this way, the HHG spectra for $a_0>0.3$ can be reproduced from the model by adjusting the value of $\delta r_{\rm A1}$, as shown by Fig.~3(b). Setting $\delta r_{\rm A1} = 0.25\lambda_0$ and $0.4\lambda_0$ recovers the HHG spectra from PIC simulations with $a_0 = 0.5$ and $1$ (adjusted to the fifth harmonic), respectively. Notably, the power-law exponent depends very sensitively on the amplitude, therefore most of the harmonic signal comes from A1 when SWB occurs.

%In this way, the ROW model can be extended to ultra-relativistic intensities as shown by Fig.~3(b).
%
%which allows half of the window to oscillate with a larger amplitude.
%the HHG spectra for $a_0>0.3$ can be reproduced from the model by adjusting the value of $\delta r_{\rm A1}$. Setting $\delta r_{\rm A1} = 0.25\lambda_0$ and $0.4\lambda_0$ recovers the HHG spectra from PIC simulations with $a_0 = 0.5$ and $1$ (adjusted to the fifth harmonic), respectively.

The harmonic generation is enhanced dramatically by the SWB effect. In particular, the PIC simulations [Fig.~3(b)] suggest the
power-law scaling is the same ($I_n\propto n^{-3.5}$) for a sufficiently strong ($a_0>2$) CP laser beam diffracting at close-to-normal incidence, which agrees well with the prediction from our model for $\delta r_{\rm A1} = 0.5\lambda_0$. This suggests the detailed electron
dynamics at the periphery is not crucial for the HHG scaling; it is sufficient to consider a sinusoidal oscillation with an amplitude limited by causality.
Because the electron layer can only travel inwards for less than half a laser cycle, the maximum displacement of the rim is $\sim c\times0.5T_0 = 0.5\lambda_0$.
%This is valid as long as the electrons on the rim oscillate predominately in the transverse plane.
In addition, both the model and simulations suggest this limit changes little with varying the aperture radius, two examples are given in Fig.~3(c).
%, i.e. a  relativistic CP laser diffracting through an aperture at close-to-normal incidence

Note that the drop-off observed at eighth and ninth harmonics on the spectra is due to limited numerical resolution. We show with higher-resolution 2D simulations that the $I_n\propto n^{-3.5}$ scaling is retained to much higher harmonic numbers without significant drop-off (see \textbf{Supplemental Material}).

Finally, the HHG fields can be obtained by filtering the diffracted fields calculated from Eq.~(1) within a certain frequency range. Using the same parameters as in Fig.~1, and setting $\delta r_{\rm A1} = 0.5\lambda_0$, the corresponding second, third, and fourth harmonics are presented in Figs.~3(d-f), respectively. Apparently the results confirm the relation $l = (n-1)\sigma$, and the harmonic fields agree very well with the PIC simulations shown in Figs.~1(g-i).\\

In conclusion, we have demonstrated that high harmonics are generated when a high-power CP laser pulse diffracts through a small aperture on a thin foil. In this process, the SAM of the driving laser beam is converted into OAM of the harmonics, giving rise to intense circularly polarised XUV vortices, with topological charge $l = (n-1)\sigma$ for the $n$th harmonic.
By means of PIC simulation and semi-analytical modeling, we show that the harmonic spectrum is $I_n\propto n^{-3.5}$, which does not
depend very much on the driving laser intensity, provided that $a_0>2$. It would be interesting to examine this scaling at very large $a_0$, as for the
ROM mechanism~\cite{Edwards2020}, this is left for future work.\\
%A semi-analytical model is derived, which shows the high-harmonic generation and spin-orbit interaction stem from the chiral electron oscillation on the rim of the aperture, that act as a ``relativistic oscillating window".\\

\begin{acknowledgments}
The author acknowledges fruitful discussions with A. Pukhov, K. Hu, T. F\"ul\"op, and I. Pusztai. This work is supported by the Olle Engqvist Foundation, the Knut and Alice Wallenberg Foundation and the European Research Council (ERC-2014-CoG Grant No. 647121). Simulations were performed on resources at Chalmers Centre for Computational Science and Engineering (C3SE) provided by the Swedish National Infrastructure for Computing (SNIC).
\end{acknowledgments}

\section{Supplemental Material}

\subsection{I.~ROW with an obliquely-incident drive laser}

\begin{figure}[!t]
\centering
\includegraphics[width=8.5cm]{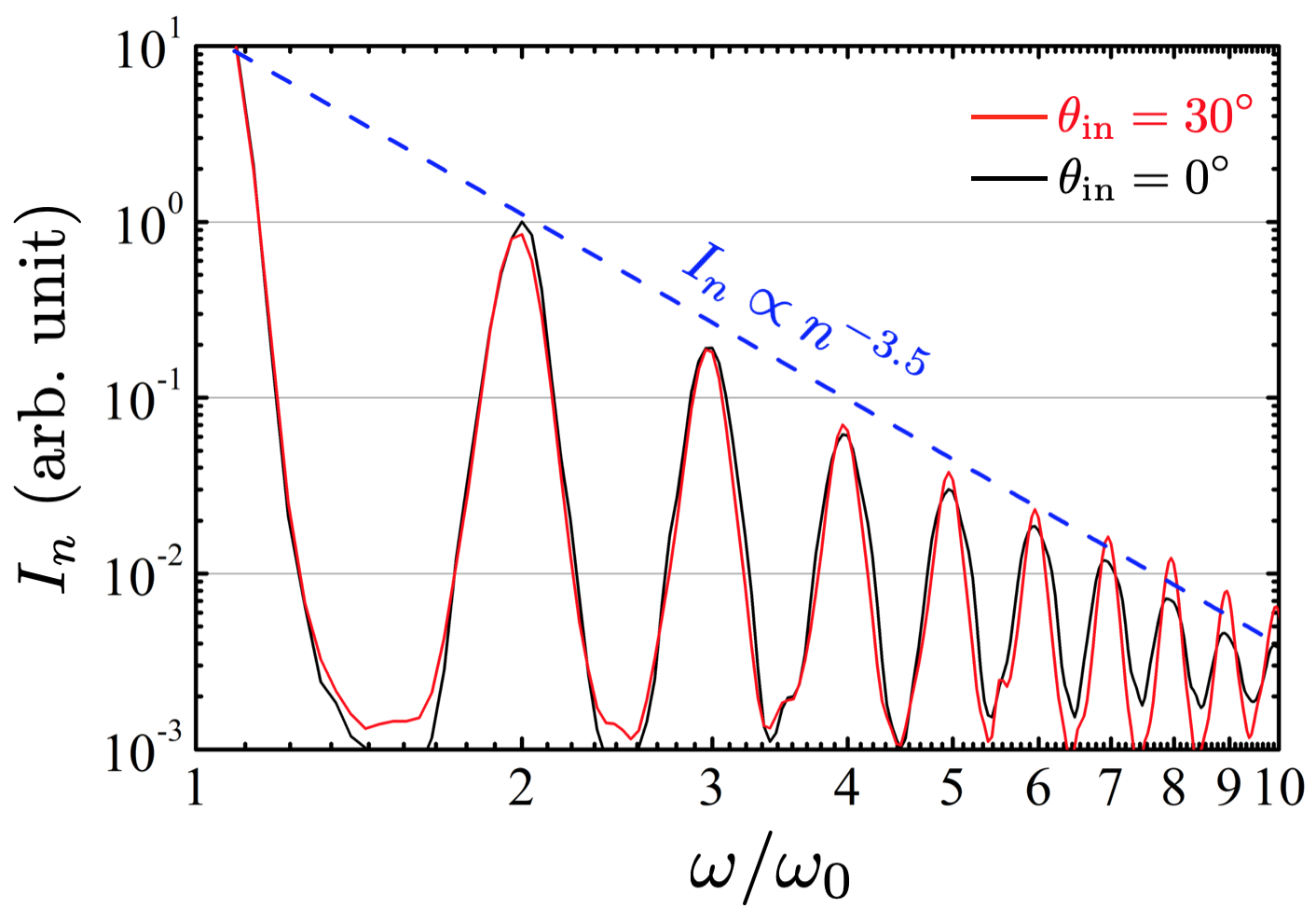}
\caption{Illustration of the effect of oblique incidence on the ROW scheme. The black and red full lines show the HHG spectra produced by a drive laser irradiating at normal incidence and 30$^{\circ}$ incident angle, respectively. The blue dashed line shows the $I_n\propto n^{-3.5}$ scaling predicted by the theory.}
\vspace{-1pt}
\end{figure}

Figure~4 shows the comparison of the spectra of harmonics between a normally incident laser (black curve) and an obliquely incident laser with 30$^{\circ}$ angle (red curve). The blue dashed line indicates the $I_n\propto n^{-3.5}$ scaling suggested by our model. High-harmonic beams are produced in both cases, and the scalings are very similar, supporting that the underlying physics is the same.

The laser field is prescribed as $\mathbf{E}_l = (\mathbf{e_y}+i\sigma\mathbf{e_z})E_0\exp(-r^2/w_0^2)\sin^2(\pi t/\tau_0)\exp(ik_0x-i\omega_0 t)$, where $0<t<\tau_0 = 54$ fs, $E_0 = 16$ TV/m is the laser amplitude ($a_0 = 5$), and $w_0 = 3.5~\mu$m is the size of focal spot.  Two incident angles $\theta_{\rm in} = 0^{\circ}$ and $30^{\circ}$ are considered.
The plasma parameters are the same as Fig.~1: the electron density is $n_0 = 30n_{\rm{c}}$ everywhere except inside the aperture, the thickness of the foil is $L_{\rm f} = 0.25~{\rm \mu m}$. The radius of the aperture is $r_{\rm{A}} = 4.0~{\rm \mu m}$, with a density gradient at the inner boundary $n(r) = n_0\exp[(r-r_{\rm{A}})/h]$ for $r<r_{\rm{A}}$, where $h = 0.2 {\rm \mu m}$ is the scale length.

Note that a Gaussian laser beam is used here, and we apply open boundary conditions to all the simulation walls. The simulation resolution is $\Delta x = \lambda_0/200$,  $\Delta y = \Delta z = \lambda_0/30$, higher than that used in Fig.~1 ($\Delta x = \lambda_0/160$,  $\Delta y = \Delta z = \lambda_0/20$), in order to check numerical convergence. The size of the simulation box and the number of macroparticles per cell are the same as that in Fig.~1.

An incident angle up to $30^{\circ}$ makes only slight difference to the HHG scaling; the small difference is mainly caused by the longitudinal oscillation of the electrons on the rim (in addition to their chiral transverse oscillation). However, according to Chen {\it et al.} \cite{Chen2016}, the longitudinal oscillation driven by a CP laser is negligible for incident angles smaller than 45$^{\circ}$. Therefore we conclude that a misalignment error in experiments does not change our results.

\subsection{II.~ROW with a self-generated aperture}

When a thin foil is irradiated by an intense laser, a relativistic plasma aperture (self-generated aperture) \cite{GI2016NP} is generated. This can be an alternative way to implement the ROW mechanism; as the aperture and the laser field are then naturally aligned, no alignment issue could possibly arise in such an experiment.

\begin{figure*}[!t]
\centering
\includegraphics[width=0.85\textwidth]{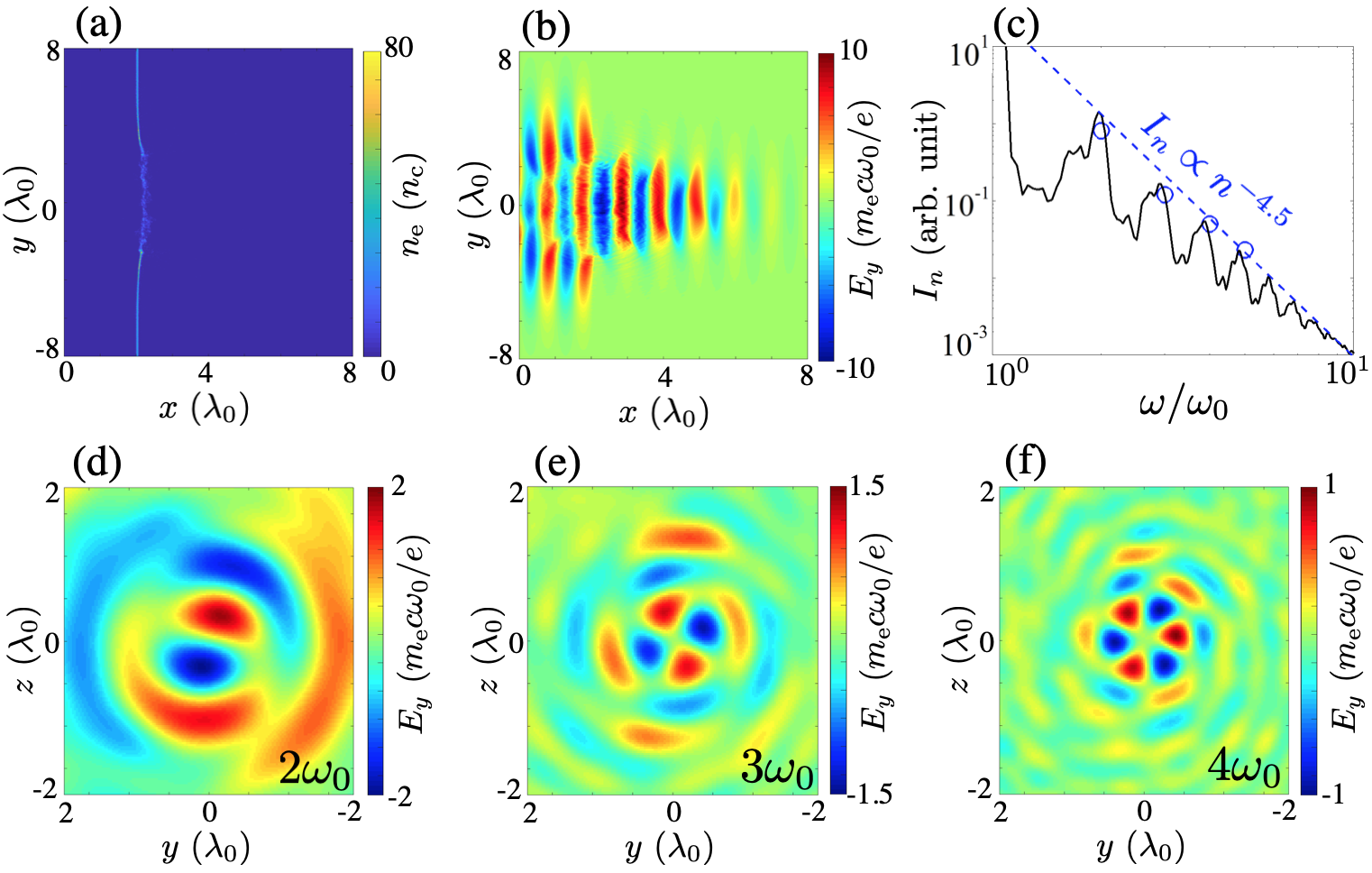}
\caption{High-harmonic generation by an intense laser diffracting through a self-generated aperture. The electron density (a) and the $E_y$ field (b) are presented in the $x$-$y$ plane, when the peak of the laser pulse  arrives at the thin foil. The spectrum observed at the rear of the thin foil (c) indicates that harmonics are being generated in the diffracted light. Panels (d-f) show the harmonic fields in the transverse ($y$-$z$) plane with frequency $2\omega_0$, $3\omega_0$, and $4\omega_0$, respectively. The blue dashed line in (c) represents a power-law fit with $I_n\propto n^{-4.5}$, the blue circles show the prediction from our model for $a_0 = 1$.}
\vspace{-1pt}
\end{figure*}

The results are presented in Fig.~5. Panels~(a-b) show the electron density and the $E_y$ field in the $x$-$y$ plane at the simulation time $t = 12 T_0$ (when the peak of the laser pulse reaches the foil). One can see that the target is destroyed by the radiation pressure and the laser is travelling through. High-order harmonics are being generated as shown by the spectrum of the transmitted light [Fig.~5(c)]. Figure~5(d-e) show the second, third, and fourth harmonic fields ($y$-component) in the transverse cross section ($y$-$z$ plane), respectively. These are vortex beams with topological number $l = (n-1)\sigma$, which show little differences to Fig.~1(g-i).

The default simulation parameters are the same as in Fig.~4, with the following changes: the  normalised laser amplitude $a_0 = 10$ and the thickness of the foil is $L_{\rm f} =60$ nm, with no pre-drilled apertures. The resolution is $\Delta x = \lambda_0/100$, $\Delta y = \Delta z = \lambda_0/40$.

The main difference with a self-generated aperture is the power-law scaling $I_n \propto n^{-4.5}$; the exponent is smaller than the one predicted in the ultra-relativistic limit ($I_n \propto n^{-3.5}$).
However, this is expected, as the laser amplitude at the edge of a self-generated aperture is much weaker than its peak amplitude -- a stronger laser drills a larger hole on the foil, but the intensity on the edge remains similar. Since the plasma density within a self-generated aperture is near-critical, one expects the laser $a_0$ acting on the edge to be close to unity. In fact, the HHG scaling shown in Fig.~5(c) is very similar to the cases with pre-drilled apertures and laser $a_0 \sim 1$ [blue circles, also presented in Fig.~3(b)].

\subsection{III.~Cut-off of the HHG spectra}

A detailed study regarding the cut-off of the spectrum is outside the scope of the current work, as it can not be addressed within the framework of the semi-analytical theory introduced in this work. This question is therefore left for future research.

\begin{figure}[!t]
\centering
\includegraphics[width=8.5cm]{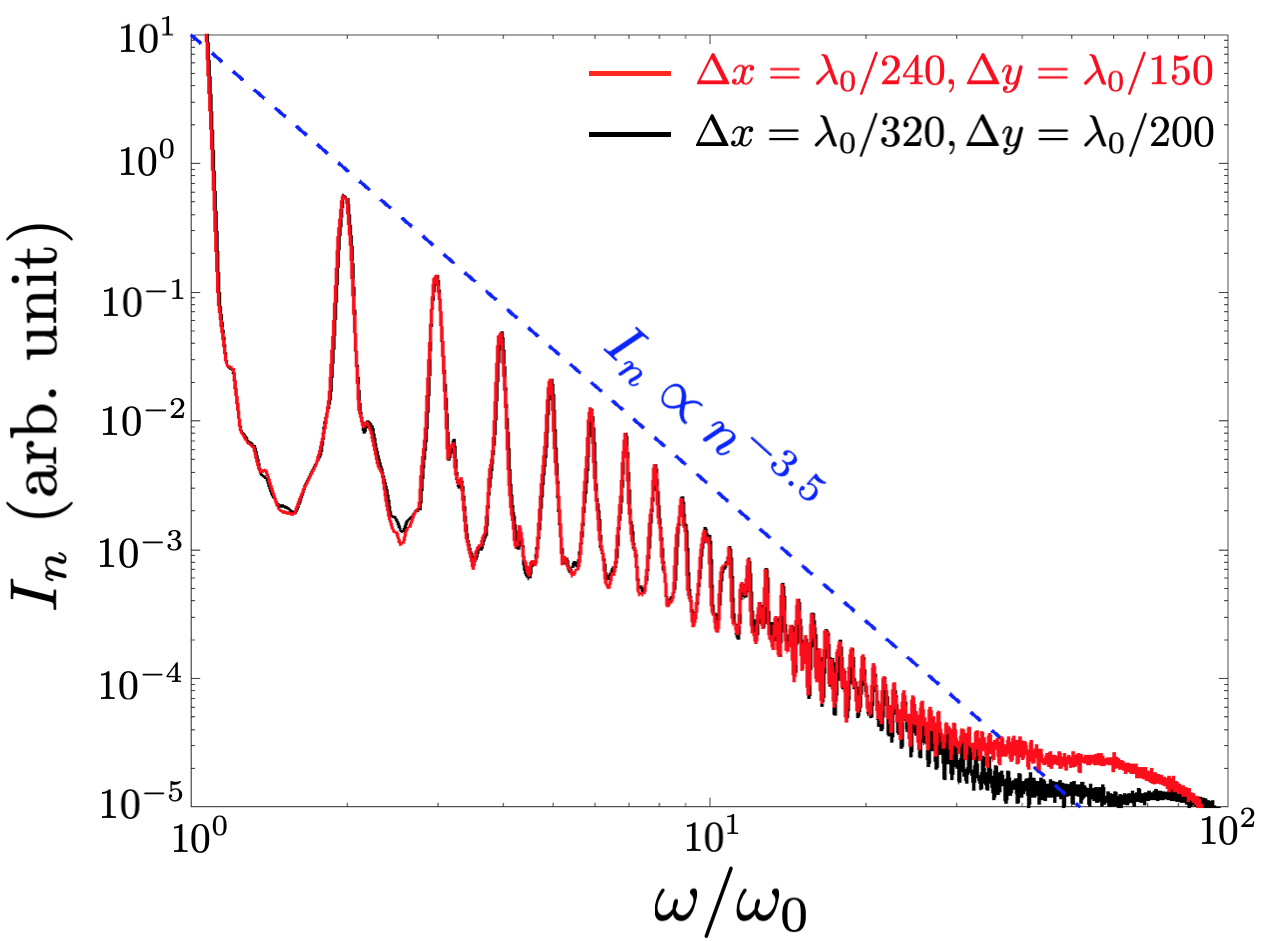}
\caption{The spectrum of high-order harmonics obtained in two-dimensional PIC simulations. The red and black full lines show the results from two simulations with different resolution to check numerical convergence. The blue dashed line shows the $I_n\propto n^{-3.5}$ scaling predicted by the theory.}
\vspace{-1pt}
\end{figure}

Here, we would like to point out the drop-off observed at the eighth and ninth harmonics on Fig.~1(c) and Figs.~3(b-c) is due to limited numerical resolution. This can be confirmed by 2D simulations with much finer resolution. We have also checked the numerical convergence with different resolutions. The results are shown in Fig.~6.

The default simulation parameters are the same as the 3D simulation presented in Fig.~1, with following changes: the normalised laser amplitude is $a_0 = 10$.
The size of the 2D simulation box is $L_x\times L_y = 15 \mu$m$\times 16 \mu$m. Two sets of numerical resolution is used to check the numerical convergence: (1) $\Delta x = \lambda_0/240$, $\Delta y = \lambda_0/150$ ($N_x\times N_y = 3200\times2400$); and (2) $\Delta x = \lambda_0/320$, $\Delta y = \lambda_0/200$ ($N_x\times N_y = 4800\times3200$). The number of macro particles in each cell are 5 for C$^{6+}$, 5 for H$^{+}$, and 35 for electrons.

It should be noted that the diffraction ``aperture" in 2D simulations is in fact a slit. However, the underlying physical processes, i.e. Doppler effect between a moving source (diffraction aperture/slit shaken by the laser) and a fixed observer, as well as the surface wave breaking effect that leads to an enhancement of the oscillating amplitude in half of the cycle, are the same as in the 3D simulations.

As one can see, the numerical convergence can be confirmed up to a harmonic order of $n \approx 25$, and the $I_n\propto n^{-3.5}$ scaling is retained to (at least) the 25th harmonic without significant drop-off.\\

\end{document}